\documentclass[prd,nofootinbib,floats,twocolumn,superscriptaddress]{revtex4-1}
\usepackage{slashed,amsmath,amsfonts,amssymb,hyperref,graphicx,bm}

\newcommand{\eq}[1]{(\ref{#1})}
\newcommand{\beq}{\begin{equation}}
\newcommand{\eeq}{\end{equation}}
\newcommand{\beqn}{\begin{eqnarray}}
\newcommand{\eeqn}{\end{eqnarray}}
\newcommand{\beqs}{\begin{subequations}}
\newcommand{\eeqs}{\end{subequations}\\[-2mm]\noindent}
\newcommand{\pr}{^{\prime}}

\newcommand{\cZ}{{\mathcal{Z}}}
\newcommand{\cL}{{\mathcal{L}}}

\renewcommand{\P}{{\cal{P}}}
\newcommand{\T}{{\cal{T}}}
\newcommand{\PT}{{\mathcal{PT}}}
\newcommand{\NJL}{\mathrm{NJL}}
\newcommand{\NH}{\mathrm{NH}}
\newcommand{\avr}[1]{\left\langle #1 \right\rangle}
\newcommand{\dirac}{\slashed\partial}
\newcommand{\bs}{\boldsymbol}

\newcommand{\R}{{\mathbb R}}
\def\bbbone{{\mathchoice {\rm 1\mskip-4mu l} {\rm 1\mskip-4mu l} {\rm 1\mskip-4.5mu l} {\rm 1\mskip-5mu l}}}

\begin{document}
\title{Spontaneous Non-Hermiticity in Nambu--Jona-Lasinio model}

\author{M.N. Chernodub}
\affiliation{Institut Denis Poisson UMR 7013, Universit\'e de Tours, 37200 France}
\affiliation{Pacific Quantum Center, Far Eastern Federal University, Sukhanova 8, Vladivostok 690950, Russia}
\author{Alberto Cortijo}
\affiliation{Departamento de F\'isica de la Materia Condensada,
Universidad Aut\'onoma de Madrid, Madrid E-28049, Spain}
\author{Marco Ruggieri}
\affiliation{School of Nuclear Science and Technology, Lanzhou University, 222 South Tianshui Road, Lanzhou 730000, China}

\date{\today}

\begin{abstract}
We explore the physical consequences of a scenario when the standard Hermitian Nambu--Jona-Lasinio (NJL) model spontaneously develops a non-Hermitian $\PT$-symmetric ground state via dynamical generation of an anti-Hermitian Yukawa coupling. We demonstrate the emergence of a noncompact non-Hermitian (NH) symmetry group which characterizes the NH ground state. We show that the NH group is spontaneously broken both in weak- and strong-coupling regimes. In the chiral limit at strong coupling, the NH ground state develops inhomogeneity, which breaks the translational symmetry.  At weak coupling, the NH ground state is a spatially uniform state, which lies at the boundary between the $\PT$-symmetric and $\PT$-broken phases. Outside the chiral limit, the minimal NJL model does not possess a stable non-Hermitian ground state. 
\end{abstract}

\maketitle

\section{Introduction}

The Nambu--Jona-Lasinio (NJL) model~\cite{Nambu:1961tp} describes the dynamics of interacting relativistic fermions. The model is often employed as a viable low-energy effective theory of quantum chromodynamics (QCD) because the NJL model, similarly to QCD, exhibits both the dynamical mass gap generation and the axial (chiral) symmetry breaking. The model naturally describes basic chiral features of the QCD ground state as well as particularities of the mesonic spectrum, and captures effects of finite temperature and baryonic density~\cite{Klevansky:1992qe}.

Similarly to QCD, the NJL model is described by a Hermitian Lagrangian. The ground states of both theories are naturally assumed to be Hermitian. In our definition, the ``Hermiticity'' of the ground state means that elementary excitations over this ground state possess a Hermitian dynamics and, therefore, are described by a Hermitian model. In our paper, we attempt to describe a scenario, when a Hermitian model may spontaneously, via a dynamical mechanism, generate a non-Hermitian ground state. 

Non-Hermitian terms usually appear in open systems when a physical system interacts an the external environment. Due to the energy exchange with the environment, some of those systems reside in an off-equilibrium but steady regime. A well-balanced steady state may be described by a Hamiltonian with real-valued eigenenergies. This is the case of the class of Parity-Time ($\PT$) symmetric systems, where the Hermiticity condition is traded by the commutation of the Hamiltonian (and the eigenstates) with the combined parity $\P$ and time reversal inversion $\T$ operation~\cite{Bender:2005}. This combined symmetry allows for an unitary dynamical evolution and real spectrum. As it was emphasized in Ref.~\cite{Mannheim:2015hto}, the crucial feature of the $\PT$ symmetry lies in the anti-linearity of the time-reversal transformation $\T$. The anti-linearity is more important for the self-consistent description of stable quantum-mechanical and field-theoretical systems than the Hermiticity of the Hamiltonian itself.

Non-Hermitian physics has previously been addressed in the contexts of interacting models of relativistic fermions. For example, the inclusion of non-Hermitian $\PT$-symmetric interactions may support real energy spectra in fermionic theories in 1+1 and 3+1 dimensions~\cite{Beygi:2019qab}. Anti-Hermitian Yukawa interactions may lead to an anomalous radiative mass-gap generation in a model of the right-handed sterile neutrinos, while the energetics of the system forbids the emergence of a dynamically generated  mass~\cite{Alexandre:2020bet,Mavromatos:2020hfy}. Inclusion of a particular $\PT$- and chirally symmetric bilinear non-Hermitian term contributes to the mass gap generation in the NJL model and leads to a rich phase structure~\cite{Felski:2020vrm}. 

Our proposal differs from the already existing approaches. Instead of incorporation of a non-Hermitian coupling to the original model,  we consider the possibility that a perfectly Hermitian model develops, via a dynamical mechanism, a non-Hermitian ground state with a physically meaningful set of features. In more detail, we look for a non-Hermitian ground state which differs from the standard Hermitian ground state in the structure of its non-perturbative vacuum condensates. We explore the possibility that the vacuum develops particular condensates that make the (quasi)particle excitations non-Hermitian. In the context of the NJL model, the fermion (quark) excitations over the non-Hermitian ground state possess non-Hermitian masses, hence the name ``non-Hermitian vacuum''.

Non-Hermitian physics of relativistic fermions may appear in fireballs of quark-gluon plasma (QGP) created in 
heavy-ion collisions. A fireball of QGP is a relativistically expanding out-of-equilibrium system. Although this system does not reside in a steady regime, the fermionic interactions may generate a non-Hermitian ground state in a steady-state non-equilibrium regime, which is realized in between the early moments of the plasma until the evolution of the QGP approaches the chiral crossover and eventual hadronization. The non-Hermitian description of the QGP in the expanding non-equilibrium regime is supported by the widely acknowledged fact that the equilibrium ground state of a generic non-Hermitian Hamiltonian is often related to an out-of-equilibrium ground state of an appropriate Hermitian system~\cite{review:1}. In many cases, the Hermitian and non-Hermitian Hamiltonians are indeed related to each other by a (non-Unitary) similarity transformations. In a coherence with the above remarks, we notice that non-Hermitian Dirac fermions allow for realization of the chiral magnetic effect~\cite{Fukushima:2008xe} in the thermal equilibrium regime~\cite{Chernodub:2019ggz}, while the similar effect is forbidden in the thermal equilibrium for the ordinary Hermitian Dirac fermions~\cite{Rubakov:2010qi,Gynther:2010ed,Landsteiner:2012kd,VF13,Yamamoto:2015fxa,Zubkov:2016tcp}. The concept of non-Hermitian quantum theory may also be extended to gauge-gravity duality~\cite{Arean:2019pom}.

In Condensed Matter Physics, the situation is similar. The problem of symmetry-broken states in interacting many-body systems with already incorporated non-Hermiticities has been studied in several physical systems in the recent years. The questions of interest include the effect of non-Hermitian terms in topological superconductivity~\cite{Kawabata:2018}, how the phenomenology attributed to Majorana states might appear in the topologically trivial region due to coupling to the environment~\cite{Avila:2019}, or the possibility of non-Hermitian superfluidity with a complex-valued interaction constant~\cite{Yamamoto:2019}. Also some problems known to reside in the strongly coupling regime have been investigated when non-Hermitian terms are considered. These are the case, for example, of the Kondo effect~\cite{Lourenso:2018}, the out-of-equilibrium-induced coupling between the Higgs mode and the Leggett modes in driven superconductors~\cite{Krull:2016}, or the Kibble-Zurek mechanism in non-Hermitian environments~\cite{Dora:2019}. In all these cases, the non-Hermitian elements are associated to open systems coupled to the environment or when the system are driven out of equilibrium. Only recently, the possibility of a non-Hermitian superconducting ground states out of an interacting Hermitian system has been proposed~\cite{Ghatak:2018}. The non-Hermitian albeit $\PT$-symmetric superconductivity displays a different phenomenology than the Hermitian counterpart. 

Before explaining the organization of the present manuscript, we want to summarize the main results the reader will find in the remaining paragraphs:
\begin{itemize}
    \item It is possible to obtain NH ground state solutions from an Hermitian NJL theory.
    \item The axial symmetry breakdown pattern that characterized the NJL model strongly changes as the NH bosonized theory does not possesses this symmetry. This statement implies strong changes in the way to physically interpret the condensates.
    \item We show that in the NH ground state, non-homogeneous solutions are energetically preferable than homogeneous ones. 
\end{itemize}
These claims go beyond the behavior of the non-Hermitian theory with respect to the $\mathcal{PT}$ symmetry.

The organization of this article is as follows: In Section~\ref{sec:free:fermions} we briefly review the phenomenology of a free model of non-Hermitian Dirac fermions where the non-Hermiticity is explicitly imposed at the Lagrangian level. The NJL model and its standard Hermitian ground state are introduced in Section~\ref{sec:NJL:H}. The homogeneous non-Hermitian ground states are discussed in Section~\ref{eq:NJL:NH} while their inhomogeneities are address within the Ginzburg-Landau approach in Section~\ref{sec:GL}. An outlook on further development and our conclusions are given in the last two sections. The Appendix contains the presentation of the gradient expansion of the effective non-Hermitian action with the nonhomogeneous terms included.

\section{Free non-Hermitian fermions}
\label{sec:free:fermions}

A possible consistent $\PT$-invariant non-Hermitian extension of the Lagrangian for a Dirac fermion~\cite{Bender:2005hf,Alexandre:2015kra,Alexandre:2017fpq,Alexandre:2017foi,Alexandre:2017erl} has the following form:
\beqn
\cL_\psi = \bar\psi \left(i \dirac - m - m_5 \gamma^5 \right) \psi,
\label{eq:cL}
\eeqn
where $m$ is the mass of the fermion field $\psi$ and $m_5$ is a non-Hermitian mass. Standard notations for the fermionic fields will be employed throughout the article: $\bar\psi = \psi^\dagger \gamma^0$, and $\dirac = \gamma^\mu \partial_\mu$ where $\gamma^\mu$ and $\gamma^5$ are the Dirac matrices.

In the momentum space, the Hamiltonian of the system~\eq{eq:cL} is given by the following operator,
\beqn
{\hat H} = {\bs \alpha} \cdot {\bs k} + \beta m + \beta \gamma_5 m_5,
\label{eq:H}
\eeqn
where ${\bs \alpha} = \gamma^0 {\bs \gamma}$ and $\beta = \gamma^0$ are the original Dirac notations. Due to the presence of the last term, the Hamiltonian~\eq{eq:H} is not a Hermitian operator: ${\hat H} \neq {\hat H}^\dagger$ provided $m_5 \neq 0$.

The Hamiltonian~\eq{eq:H}, however, commutes with the product of parity $\P$ and time $\T$ inversions: $[\PT,{\hat H}] = 0$, thus implying that the theory respects the $\PT$ symmetry. These discrete operations affect the coordinates as follows: $\P (t, {\bs x}) = (t, - {\bs x})$ and $\T (t, {\bs x}) = (- t, {\bs x})$. The parity inversion $\P$ acts linearly on the fermionic fields:
\beqn
\P:\quad 
\left\{
\begin{array}{rcl}
\psi(t,{\bs x}) & \to & \gamma^0 \psi(t,{\bs x}), 
\\[2mm]
{\bar\psi}(t,{\bs x}) & \to & {\bar \psi}(t,{\bs x}) \gamma^0, 
\end{array}
\right.
\eeqn
while the time inversion $\T$ is represented by an anti-linear operator:
\beqn
\T:\quad 
\left\{
\begin{array}{rcl}
\psi(t,{\bs x}) & \to & i \gamma^1 \gamma^2 \psi^*(t,{\bs x}), 
\\[2mm]
{\bar\psi}(t,{\bs x}) & \to & {\bar \psi}^*(t,{\bs x}) i \gamma^1 \gamma^2.
\end{array}
\right.
\eeqn
Therefore, the Lagrangian~\eq{eq:cL} describes a $\PT$ symmetric non-Hermitian theory of Dirac fermions.

The Lagrangian~\eq{eq:cL} gives the following classical equations of motion:
\beqn
\left(i \dirac - m - m_5 \gamma^5 \right) \psi & = & 0\,, \\
\bar{\psi}\left(i \overleftarrow{\dirac} + m + m_5 \gamma^5 \right) & = & 0\,.
\eeqn
The positive-frequency solutions of the Dirac equation,
\beqn
\psi(x) = u(p) e^{- i p\cdot x},
\eeqn
are expressed via the spinor $u(p)$ which satisfies the Dirac equation in the momentum space:
\beqn
\left({\slashed p} - m - m_5 \gamma^5 \right) u(p) = 0,
\label{eq:Dirac:p}
\eeqn
$p^\mu = (p^0,{\bs p})$ is the the four-momentum  and $p\cdot x = p_\mu x^\mu$ is the scalar product in the Minkowski space. 
The self-consistency of Eq.~\eq{eq:Dirac:p} requires $p^2 = M^2$ and determines the energy spectrum ($p_0 \equiv E_{\bs p}$):
\beqn
E_{\bs p} = \sqrt{{\bs p}^2 + M^2},
\eeqn
where the mass of the fermionic excitation is
\beqn
M = \sqrt{m^2 - m_5^2}.
\label{eq:M:0}
\eeqn
The same statement applies also to the negative-energy solutions.
We always take $|m| \geqslant |m_5|$ to insure that the mass $M$ is a real quantity. This range of parameters correspond to the ``$\PT$-symmetric'' phase. If the Hermitian mass $|m|$ is smaller than the non-Hermitian mass $|m_5|$, then the system resides in the ``$\PT$-broken'' phase which is characterized by paired complex branches of fermionic energies that make the vacuum unstable. The stability of the non-Hermitian theory in the $\PT$ symmetric region is the direct consequence of $\PT$ symmetry of the ground state solution of the  Hamiltonian~\eq{eq:H}.

In the absence of both non-Hermitian $m_5$ and Hermitian $m_5$ masses, the free fermionic theory~\eq{eq:cL} possesses the global vector and global axial symmetries described by the continuous transformations, respectively:
\beqs
\beqn
U(1)_V: \qquad & \psi & \to e^{i \omega_V} \psi,
\qquad
{\bar \psi} \to {\bar \psi} e^{- i \omega_V},
\label{eq:U1:V}\\
U(1)_A: \qquad & \psi & \to e^{i \gamma^5 \omega_A} \psi,
\quad\,
{\bar \psi} \to {\bar \psi} e^{i \gamma^5 \omega_A},
\label{eq:U1:A}
\eeqn
\label{eq:symmetries}
\eeqs
with the coordinate-independent parameters $\omega_V$ and $\omega_A$.

In the limit of the vanishing masses, $m = m_5 = 0$, the continuous symmetries~\eq{eq:symmetries} lead to the following vector and pseudo-vector Noether currents, respectively:
\beqs
\beqn
j^\mu_V & = & \bar\psi \gamma^\mu \psi\,,
\label{eq:j} \\
j^\mu_A & = & \bar\psi \gamma^\mu \gamma^5 \psi\,,
\label{eq:j5}
\eeqn
\label{eq:jj5}
\eeqs
which are conserved at the classical level.

However, if the masses $m$ and $m_5$ are nonzero, then the currents~\eq{eq:jj5} are no more conserved:
\beqs
\beqn
\partial_\mu j^\mu_V & = & - 2 i m_5 \bar\psi \gamma^5 \psi\,, 
\label{eq:djmu}\\
\partial_\mu j^\mu_A & = & \phantom{-}2 i m \,\bar\psi \gamma^5 \psi\,.
\label{eq:dj5mu}
\eeqn
\label{eq:djj5mu}
\eeqs
The Hermitian mass $m$ breaks the axial $U(1)_A$ symmetry~\eq{eq:U1:A}
while the non-Hermitian mass $m_5$ breaks the vector $U(1)_V$ symmetry~\eq{eq:U1:V}.

However, one can see from the non-conservation pattern~\eq{eq:djj5mu} that the model admits the following linear combination of the currents~\eq{eq:jj5}:
\beqn
J^\mu_V = j^\mu_V + \frac{m_5}{m} j^\mu_A,
\label{eq:J:V}
\eeqn
which is conserved at the classical level:
\beqn
\partial_\mu J^\mu_V = 0.
\label{eq:conservation}
\eeqn
The definition~\eq{eq:J:V} gives us the new conserved vector current. We call the expression in Eq.~\eq{eq:J:V} a current because it follows a conservation law, Eq.~\eq{eq:conservation}.

The new axial current is orthogonal to the new vector current~\eq{eq:J:V}:
\beqn
J^\mu_A = j^\mu_A - \frac{m_5}{m} j^\mu_V.
\label{eq:J:A}
\eeqn
This quantity is not conserved at the classical level:
\beqn
\partial_\mu J^\mu_A = 2 i \frac{m^2 - m_5^2}{m} \,\bar\psi \gamma^5 \psi,
\eeqn
similarly to the ordinary axial current. 

A system with a finite density of the conserved charge $J^0_V$ may be controlled by the thermodynamically conjugated chemical potential. 
Certain consequences of the finite-density non-Hermitian fermions are discussed in Ref.~\cite{Chernodub:2019ggz}. Below we proceed to the investigation of the ground state of a Hermitian NJL model in which the non-Hermitian theory~\eq{eq:cL} may arise as an effective theory in a spontaneously formed non-Hermitian ground state.

\section{Hermitian ground state in the NJL model}
\label{sec:NJL:H}

\subsection{Nambu-Jona--Lasinio model}

The NJL model provides a simplest description of interacting fermions which features the axial (chiral) symmetry breaking and the mass gap generation. We consider the standard Hermitian NJL model described by the following Lagrangian:
\begin{equation}
\cL_{\NJL} = \bar{\psi} (i \slashed\partial - m_0)\psi +\frac{G}{2}\left[\left(\bar{\psi}\psi\right)^2+ \left(\bar{\psi}i\gamma_5 \psi \right)^2 \right]\,,
\label{eq:L:NJL:1}
\end{equation}
which describes the single fermion species $\psi$ with a bare mass $m_0$. The coupling constant $G$ determines the strength of the local four-fermion interaction. For a single-species model that we consider in our manuscript, the terms axial and chiral can be used interchangeably. 

In the massless limit, $m_0 = 0$, the NJL Lagrangian~\eq{eq:L:NJL:1} is invariant under vector~\eq{eq:U1:V} and axial~\eq{eq:U1:A} global transformations. The $U(1)_V$ symmetry may be gauged via the coupling of the fermions to the electromagnetic field. Since in the physical world the vector (electric) current is always conserved, the $U(1)_V$ symmetry should never be broken in realistic models. However, the $U(1)_A$ may become broken at the quantum level. In the single-species model that we consider~\eq{eq:L:NJL:1}, the breaking of the chiral symmetry appears spontaneously due to the four-fermion interaction as we discuss below.\footnote{In the single-fermion model, the $U(1)_A$ symmetry~\eq{eq:U1:A} may be called both axial and chiral. This terminological degeneracy is lifted in the multi-species models.}

In the presence of the fermionic mass, $m_0 \neq 0$, the $U(1)_V$ symmetry~\eq{eq:U1:V} is maintained, while the axial symmetry~\eq{eq:U1:A} gets explicitly broken. In QCD, however, the bare mass (also called ``current mass'') $m_0$ is much smaller then the dynamically generated fermionic mass. Therefore, the effects of the axial mass are small and the axial symmetry~\eq{eq:U1:A} is said to be approximately correct. 
Below, we review the derivation of the standard Hermitian ground states and the dynamical mass gap generation in the NJL model~\eq{eq:L:NJL:1}. Afterwards, we will proceed to the investigation of the non-Hermitian mass gap generation in the same model. 

In the scenario we are considering here, the non-Hermiticity is broken dynamically within the Hermitian model. For a different version of the NJL model, where the non-Hermiticity is broken explicitly via a coupling of the NJL Lagrangian~\eq{eq:L:NJL:1} to a non-Hermitian background, we refer the reader to Ref.~\cite{Felski:2020vrm}.

Before proceeding further, it is necessary to comment on the terminology that we use in our paper. Usually, one uses the terms Hermitian/non-Hermitian in application to the mass matrices to distinguish the Hermitian mass matrix, $m + i m_5 \gamma_5$ from its non-Hermitian counterpart, $m + m_5 \gamma_5$. Here we use this terminology to describe not only the matrices but also the ground states. In order to justify the terminology and clarify more the central idea of the paper, let us consider the example of a spontaneously broken theory in which masses of particles are acquired with the Higgs mechanism: the expectation value of a (multicomponent) scalar particle determines a mass term for a fermionic particle. Depending on the pattern of the symmetry breaking, the mass term in the spontaneously broken ground state could be either Hermitian or non-Hermitian hence the name of the ground state: on the Hermitian ground state, the fermions have the Hermitian mass matrices while in a non-Hermitian ground state there is at least one fermion that possesses a non-Hermitian mass matrix. The same considerations are also applicable to bosonic fields.

\subsection{Hermitian ground state in the NJL model}

In the standard approach, the partition function of the NJL model~\eq{eq:L:NJL:1},
\begin{equation}
\cZ = \int D\psi D\bar{\psi} \exp\left\{i\int d^4x \, \cL_{\NJL}({\bar\psi},\psi) \right\}\,,
\label{eq:Z:NJL:1}
\end{equation}
is partially bosonized by inserting the identities\footnote{Hereafter we omit inessential normalization factors in front of the functional integrals.}
\beqs
\beqn
1 & = & \int D\sigma \exp\biggl\{- \frac{i}{2G} \int d^4x \left(\sigma + G\bar{\psi}{\psi}\right)^2 \biggr\}\,,
\label{eq:identity:1}\\
1 & = & \int D\phi \exp\biggl\{- \frac{i}{2G} \int d^4x  \left(\phi + G\bar{\psi}i\gamma _5{\psi}\right)^2 \biggr\}\,,\quad
\label{eq:identity:2}
\eeqn
\label{eq:identity:I}
\eeqs
under the integral over the fermionic fields in Eq.~\eq{eq:Z:NJL:1}.
The fields $\sigma$ and $\phi$ are real-valued quantities.

The prefactors in the exponents of Eq.~\eq{eq:identity:I} are chosen in such a way that, after the substitution of Eqs.~\eq{eq:identity:I} to Eq.~\eq{eq:Z:NJL:1}, 
\begin{equation}
\cZ = \int D\psi D\bar{\psi} D\sigma D\phi  \exp\left\{i\int d^4x \, \cL_{\NJL}({\bar\psi},\psi;\sigma,\phi)\right\}\,,
\label{eq:Z:NJL:2}
\end{equation}
the four-fermion interaction terms of Eq.~\eq{eq:L:NJL:1} disappear:
\beqn
\cL_{\NJL}({\bar\psi},\psi;\sigma,\phi) & = & 
\bar{\psi} \left[ i \slashed\partial  - \sigma - i \gamma^5 \phi)\right]\psi \nonumber \\
& & - \frac{(\sigma - m_0)^2 + \phi^2}{2 G}. 
\label{eq:L:NJL:2}
\eeqn
The partially-bosonized NJL Lagrangian~\eq{eq:Z:NJL:2} becomes a bilinear expression in terms the original fermionic fields, coupled, via the Yukawa interactions, with the scalar $\sigma$ and pseudoscalar $\phi$ auxiliary fields.\footnote{For the model with a single fermion, we reserve the letter $\phi$ for the pseudoscalar instead of the standard notation $\pi$.} In the expression~\eq{eq:Z:NJL:2} we have shifted the $\sigma$--field, $\sigma \to \sigma - m_0$ for the sake of further convenience.

The interpretation of the auxiliary fields $\sigma$ and $\phi$ may be deduced from the saddle-point equations of the partition function~\eq{eq:Z:NJL:2}:
\beqn
\avr{\sigma} = m_0 + G \avr{\bar{\psi}\psi}, 
\qquad
\avr{\phi} = G \avr{\bar{\psi}i\gamma_5 \psi}.
\label{eq:sigma:condensate}
\eeqn
The vacuum expectation value of the condensate $\sigma$ plays a role of the mass of the fermion field. At the saddle point, the vacuum expectation values of the scalar and pseudoscalar auxiliary fields acquire contributions proportional, correspondingly, to the scalar and pseudoscalar condensates of fermionic fields~\eq{eq:sigma:condensate}. 
The emergence of the scalar fermion condensate (also called ``chiral condensate'') $\avr{\bar{\psi}\psi}$ leads to two dynamical phenomena. First, the chiral condensate is not invariant under the chiral transformations~\eq{eq:U1:A} and therefore it breaks the chiral symmetry. Second, the emergence of the chiral condensate leads to the mass gap generation~\eq{eq:sigma:condensate} as the field $\sigma$ plays a role of the fermion mass~\eq{eq:L:NJL:2}. The pseudoscalar condensate vanishes in the physical vacuum 
\beqn
\avr{\bar{\psi}i\gamma_5 \psi} = 0,
\label{eq:pseudo:condensate}
\eeqn
implying, according to Eq.~\eq{eq:sigma:condensate}, that the mean value of the field $\phi$ vanishes as well, $\avr{\phi} = 0$.

The applicability of the saddle-point approximation becomes justified for the NJL model with many fermion flavors $N_f$. In the limit $N_f \to \infty$, the saddle-point calculation becomes exact. The consistency of the physical interpretation~\eq{eq:sigma:condensate} is guaranteed by the fact that the bilinears $\bar{\psi}\psi$ and $\bar{\psi}i\gamma_5 \psi$, similarly to the auxiliary fields $\sigma$ and $\phi$, are real-valued quantities for any spinor field $\psi$.

Below, we will treat the auxiliary fields $\sigma$ and $\phi$ in the mean-field approximation. For the sake of brevity, we will use the same notations for the fields and their expectation values, $\sigma = \avr{\sigma}$ and $\phi = \avr{\phi}$. In a leading order, we assume that these fields are independent of the spacetime coordinates. Later, we will consider physical excitations over this uniform background. 

The integration over the fermionic fields in the partition function~\eq{eq:L:NJL:2} leads to the following purely scalar representation of the NJL model:
\beqn
\cZ & = & \int D\sigma D\phi\exp\biggl\{- \frac{i}{2G} \int d^4 x \left[(\sigma - m_0)^2 + \phi^2\right] \nonumber \\
& & \hskip 20mm + \mathrm{ln}\,\mathrm{det} \, \bigl[ i\slashed\partial - \left(\sigma + i\gamma_5\phi \right)\bigr] \biggr\}.
\label{eq:Z:NJL:3} 
\eeqn

In the completely bosonized representation of the NJL model~\eq{eq:Z:NJL:3}, the chiral invariance~\eq{eq:U1:A} is translated to a chiral (axial) rotation for the following combination of the  scalar and pseudoscalar fields:
\beqn
\sigma + i\gamma _5\phi \to e^{-i\gamma _5 \theta} \left(\sigma + i\gamma_5\phi \right) e^{-i\gamma_5 \theta} \equiv \tilde\sigma + i\gamma_5\tilde\phi.
\quad
\label{eq:rotation}
\eeqn
For coordinate--independent condensates $\sigma$ and $\phi$, the fermionic determinant depends only on the combination $\sigma^2 + \phi^2$ which is invariant under the chiral rotation~\eq{eq:rotation}.

It is convenient to use the rotation~\eq{eq:rotation} to turn the combination $\sigma + i \gamma^5 \phi$ into the purely scalar field with the absolute value $|\tilde \sigma| = \sqrt{\sigma^2 + \phi^2}$. Taking $\tilde \sigma$ a real positive number ($\tilde \sigma > 0$ and $\tilde\phi = 0$) and renaming back $\tilde \sigma \to \sigma$, we find that the fermionic sector of partially bosonized mean-field NJL model corresponds to a theory of massive fermions with an effective mass $m = \sigma$ and vanishing pseudoscalar field $\phi = 0$ after the chiral rotation. 

Using then the expression for the functional trace of an operator ${\hat {\cal O}}$,
\beqn
\mathrm{tr} \, {\hat {\cal O}} = \int d^4 x \int \frac{d k_0}{2 \pi} \int \frac{d^3 k}{(2 \pi)^3}\, {\cal O}_{k_0,{\bs k}},
\label{eq:integration}
\eeqn
we get for the (chirally-rotated) last term in Eq.~\eq{eq:Z:NJL:3} the following expression:
\beqn
\mathrm{ln}\,\mathrm{det} \, \bigl( i\slashed\partial - \sigma \bigr) = - i \int d^4 x \, V_{\mathrm{H}}^{\mathrm{int}}(\sigma),
\eeqn
with the potential for the field $\sigma$:
\beqn
V_{\mathrm{H}}^{\mathrm{int}}(\sigma) = i \int \frac{d k_0}{2 \pi} \int \frac{d^3 k}{(2 \pi)^3}\, \mathrm{tr} \, \mathrm{ln} \frac{{\slashed k} - \sigma}{\Lambda}.
\label{eq:V:int:1}
\eeqn
Here ${\slashed k} \equiv \gamma^\mu k_\mu$ and the (so far arbitrary) quantity of the dimension of mass $\Lambda$ is needed to maintain correct dimension. Hereafter we will ignore inessential additive constants in the potentials and actions. The subscript ``H'' in Eq.~\eq{eq:V:int:1} reminds us that we are working with the Hermitian theory.

Although it is possible to continue the derivation in Minkowski space-time, it is convenient to perform the Wick rotation to the Euclidean momentum space:
\beqn
k_0 \to i k_4 \,, \qquad \int \frac{d k_0}{2 \pi} \to i \int \frac{d k_4}{2 \pi}.
\label{eq:Wick}
\eeqn
We arrive to the potential~\eq{eq:V:int:1} given by the following expression:
\beqn
V_{\mathrm{H}}^{\mathrm{int}}(\sigma) = - 2  \int \frac{d^4 k}{(2 \pi)^4}\, \mathrm{ln} \frac{k^2 + \sigma^2}{\Lambda^2},
\label{eq:V:int:2}
\eeqn
where $d^4 k = d k_4 d^3 k$. Here we used the fact that under the Wick rotation ${\slashed k} = \gamma^0 k^0 - {\bs \gamma}\cdot {\bs k} \to - \gamma^4 k^4 - {\bs \gamma}\cdot {\bs k} = - {\slashed k}$, supplemented by the standard chain of relations:
\beqn
& & \int \frac{d^4 k}{(2 \pi)^4}\, \mathrm{tr} \, \mathrm{ln} \frac{{\slashed k} - \sigma}{\Lambda} \nonumber \\
& = & \frac{1}{2} 
\int \frac{d^4 k}{(2 \pi)^4}\, \mathrm{tr} \, \left[\mathrm{ln} \frac{{\slashed k} + \sigma}{\Lambda}
+ \mathrm{ln} \frac{- {\slashed k} + \sigma}{\Lambda} \right] \\
& = & \frac{1}{2} 
\int \frac{d^4 k}{(2 \pi)^4}\, \mathrm{tr} \, \mathrm{ln} \frac{k^2 + \sigma^2}{\Lambda^2_0} = 
2 \int \frac{d^4 k}{(2 \pi)^4}\, \mathrm{ln} \frac{k^2 + \sigma^2}{\Lambda^2}. \nonumber  
\eeqn
The effective potential for the scalar field $\sigma$ is then
\beqn
V_{\mathrm{H}}(\sigma) = \frac{(\sigma - m_0)^2}{2 G} - 2  \int \frac{d^4 k}{(2 \pi)^4}\, \mathrm{ln} \frac{k^2 + \sigma^2}{\Lambda^2}.
\label{eq:V:sigma:0}
\eeqn
The kinetic term for the $\sigma$ field is absent in the mean-field approach with the uniform $\sigma$ background.

The ground state of the model is determined by the minimization of the action with respect to the $\sigma$ field. We employ the four-momentum regularization scheme with a hard ultraviolet cutoff. The regularized potential~\eq{eq:V:sigma:0}~is:
\beqn
V_{\mathrm{H}}(\sigma) & = & \frac{(\sigma - m_0)^2}{2 G} - \frac{1}{4\pi^2} \int_0^\Lambda k^3 d k\, \mathrm{ln}\frac{k^2 + \sigma^2}{\Lambda^2} \nonumber \\
& = & \frac{\sigma^2}{2 G} + \Lambda^4 v\left(\frac{\sigma^2}{\Lambda^2}\right),
\label{eq:V}
\eeqn
where the quantity $\Lambda$ serves an ultraviolet cutoff which plays a role of a physical parameter in this model and
\beqn
v(x) = - \frac{1}{16 \pi^2} \left[x + x^2  \ln x + \left(1 - x^2\right) \ln(x + 1) \right]. \qquad
\label{eq:small:v}
\eeqn

The critical coupling,
\beqn
G_c = 4 \pi^2 \Lambda^{-2} \simeq 39.48\, \Lambda^{-2}\,,
\label{eq:Gc}
\eeqn
defines two regimes of the theory. Consider first the case of  zero bare mass $m_0 = 0$. For a weak coupling $G < G_c$, the potential $V(\sigma)$ takes its minimum at vanishing field $\sigma = 0$ which defines a chirally symmetric vacuum. At strong coupling $G > G_c$, the  minimum of the potential is reached at $\sigma \neq 0$, the chiral symmetry gets broken and the fermions acquire the mass $M = \avr{\sigma} \sim \Lambda$ via the dynamical mechanism. A small nonzero bare mass, $m_0 \neq 0$, shifts the minimum $\sigma$ to a nonzero value $\sigma \sim m_0$ at weak coupling $G$. However, the  dynamical mass generation overwhelm the bare mass, $\Lambda \gg m_0$, and the in the chirally broken phase, the mass of the fermion is given by the dynamically generated mass.

Alternatively, one can determine the ground state by solving a system of the mass-gap equations that correspond to the extremization of the effective potential with respect to the dynamical field: $\delta V(\sigma,\phi)/\delta \sigma = 0$ and $\delta V(\sigma,\phi)/\delta \phi = 0$. Restoring the pseudoscalar field $\phi$ in the effective potential~\eq{eq:V:sigma:0}, we obtain the following system of equations:
\beqs
\beqn
\sigma - m_0 - 4 G \int^\Lambda \frac{d^4 k}{(2 \pi)^4}\, \frac{\sigma}{k^2 + \sigma^2 + \phi^2} & = & 0, \\
\phi - 4 G \int^\Lambda \frac{d^4 k}{(2 \pi)^4}\, \frac{\phi}{k^2 + \sigma^2 + \phi^2} & = & 0, 
\eeqn
\label{eq:mass:gap}
\eeqs
which has the same solution that we discussed above. While both methods often give identical solutions, the direct minimization of the effective potential allows to verify that the ground state corresponds indeed to a globally stable state given by the global minimum of the effective potential.

It is convenient to consider the ground state of the model from the perspective of the statistical sum. To this end, one should perform the Wick rotation from the Minkowski spacetime to the Euclidean space. The potential $V_{\mathrm{H}}$ determines the statistical factor, $e^{- \int d^4 x V_{\mathrm{H}}(\sigma)}$, which provides us with statistical weight of the configuration $\sigma$. Therefore, the global minimum of the potential $V$ corresponds to the ground state in statistical equilibrium. In the mean-field approximation, the minimum is identified with the mass gap equations~\eq{eq:mass:gap}.

The above statement does not apply to the cases when the potential $V_{\mathrm{H}}$ is complex. Therefore it does not apply to the finite-density systems which suffer from the notorious sign problem  (although the sign problem is less severe and treatable in the mean-field analytical approaches) and, for example, to near-equilibrium states in uncompensated background electric fields that suffer from instabilities. We will see below that this statistical approach gives a sense to the non-Hermitian analogue of the mean-field potential~\eq{eq:V}.

Our derivation assumes the validity of the mean-field approach, which, in turn, relies on the applicability of the saddle-point approximation in the path- or statistical- integral. Strictly speaking, the saddle-point method works when the Gaussian exponent is large so that the main contribution to the path integral is given by the expansion of the integrand over the field fluctuations around the saddle point determined by Eq.~\eq{eq:mass:gap}. This approximation is valid in the limit of a large number of fermions (a large-$N$ limit), but it is routinely applied to a single (or few) species of fermions because the results in both cases are close to each other in practical cases~\cite{Klevansky:1992qe}. To add rigor to the use of the mean-field approach, it is enough to can consider a system of $N$ equivalent fermions which would put the fermionic determinant in Eq.~\eq{eq:Z:NJL:3} to the $N$th power and take the $N \to \infty$ limit. To keep our notations concise, we will continue to use a single fermion species keeping in mind that a more justified many-fermion system would give qualitatively the same results.

\section{Non-Hermitian ground state in the NJL model}
\label{eq:NJL:NH}

\subsection{Non-Hermitian bosonization}

In what follows we exploit the possibility that the partial bosonization, performed with the help of the standard identities~\eq{eq:identity:I}, is not the only possible choice that can be employed for the bosonization. 

In order to maintain the explicit Hermiticity of the model, we required for the fields $\sigma$ and $\phi$ to be real-valued quantities. Below, we leave the condition of the Hermiticity and consider the complex-valued bosonic fields.

To this end, we generalize the identities~\eq{eq:Z:NJL:1} using four, instead of the two ones~\eq{eq:identity:I}:
\beqs
\beqn
1 & {=} & \int D\sigma_1 \exp\biggl\{- \frac{i}{2 G_{\sigma1}} \int d^4x \left(\sigma_1 + G_{\sigma1}\bar{\psi}{\psi}\right)^2 \biggr\},
\label{eq:identity:3}\\
1 & {=} & \int D\sigma_2 \exp\biggl\{- \frac{i}{2G_{\sigma2}} \int d^4x \left(\sigma_2 + i G_{\sigma2}\bar{\psi}{\psi}\right)^2 \biggr\},
\label{eq:identity:4}\\
1 & {=} & \int D\phi_1 \exp\biggl\{- \frac{i}{2G_{\phi1}} \int d^4x  \left(\phi_1 + G_{\phi1}\bar{\psi}i\gamma_5{\psi}\right)^2 \biggr\}, \quad
\label{eq:identity:5}\\
1 & {=} & \int D\phi_2 \exp\biggl\{- \frac{i}{2G_{\phi2}} \int d^4x  \left(\phi_2 - G_{\phi2}\bar{\psi}\gamma_5{\psi}\right)^2 \biggr\}.\qquad\
\label{eq:identity:6}
\eeqn
\label{eq:identity:II}
\eeqs
These functional integrals should be understood in the standard path-integral sense, namely, as a product of simple integrals in every point of the spacetime. Each of the simple Gaussian integrals that enter Eqs.~\eq{eq:identity:II} can be evaluated to a finite value with any reasonable regularization. The integration result enters the path integration measure as an inessential normalization constant.

Due to the local nature of the expressions in under the exponents in Eq.~\eq{eq:identity:II}, one can consider each of these integrals as a product of elementary integrals at each point $x$. For simplicity, we denote $\sigma_a(x)$ or $\phi_a(x)$ as $f$ and use the notation $h$ for any of the bilinear condensates, $\bar{\psi}{\psi}$ or $\bar{\psi} \gamma_5 {\psi}$ that enter Eq.~\eq{eq:identity:II}. To regularize the integral, we add multiply each integrand by the factor $e^{ - \epsilon f^2}$, where $\epsilon > 0$ is a regulator which will be taken to zero at the end of the calculation. Then an elementary integral that enters the expressions in the right-hand-side of Eqs.~\eq{eq:identity:3} and \eq{eq:identity:5} can be written as follows:
\beqn
&   & \int\limits_{-\infty}^{ \infty} d f \exp \left\{ - \frac{i}{2 G} (f + h)^2 - \epsilon f^2\right\} \nonumber \\
& = & \int\limits_{-\infty}^{ \infty} d f \exp \left\{ - \left(\epsilon + \frac{i}{2 G} \right) f^2 - i f h - \frac{G^2 h^2}{2} \right\} \nonumber \\
& = &  \sqrt{\frac{2 \pi G}{2 \epsilon G + i}} \exp \left\{ - \epsilon  \frac{G^2 h^2}{1 - 2 i G \epsilon} \right\}\,.
\eeqn
Lifting out the normalization, $\epsilon \to 0$, we find that for any finite value of $h$ (the condensate) and $G$ (the coupling), the integral converges, regardless of the sign of the coupling $G$, to a condensate-independent value which is nothing by a normalization constant. The same statement is true for the integrals~\eq{eq:identity:4} and \eq{eq:identity:6} with the substitution $h \to i h$.

Inserting these identities into Eq.~\eq{eq:Z:NJL:1}, we find that we might cancel the four-fermion interaction terms of the NJL Lagrangian~\eq{eq:L:NJL:1} provided the parameters in Eq.~\eq{eq:identity:II} satisfy the following conditions:
\beqn
G_{\sigma1} - G_{\sigma2} = G, 
\qquad
G_{\phi1} - G_{\phi2} = G.
\label{eq:G:relations}
\eeqn

Repeating all steps as in the Hermitian counterpart, we get the new partially bosonized Lagrangian:
\beqn
& & \cL_{\NJL} = - \frac{1}{2} \left( \frac{(\sigma_1 - m_0)^2}{G_{\sigma1}} + \frac{\sigma^2_2}{G_{\sigma2}} + \frac{\phi^2_1}{G_{\phi1}} + \frac{\phi^2_2}{G_{\phi2}} \right)
\nonumber \\
& & \hskip 8mm +
\bar{\psi} \left[ i \slashed\partial - (\sigma_1 + i \sigma_2) - i \gamma^5 (\phi_1 + i\phi_2)\right]\psi,
\label{eq:L:NJL:NH:1}
\eeqn
in which we have redefined the field $\sigma_1 \to \sigma_1 - m_0$ for convenience.

Let us consider the fermionic part of the NJL Lagrangian~\eq{eq:L:NJL:NH:1}. For the state $\sigma_2 = 0$ and $\phi_2 = 0$, the theory reduces to the standard Hermitian case considered earlier: the axial freedom~\eq{eq:rotation} may be used to remove the field $\phi_1$ and we arrive to the bosonized mean-field theory, given by Eqs.~\eq{eq:V} and \eq{eq:small:v}, with the condensate $\sigma \equiv \sigma_1$.

In the rest of this section we assume, unless explicitly stated otherwise, that the bosonic fields do not depend on the coordinates. The case of the non-uniform background with space-dependent fields will be considered in the next section.

If the fields $\sigma_2$ and $\phi_2$ are taken to be nonzero, then the fermionic part of the Lagrangian~\eq{eq:L:NJL:NH:1} becomes non-Hermitian. The excitations of a non-Hermitian theory may possess a complex spectrum signaling instabilities. The physical spectrum corresponds to the poles $k_0 = \pm \omega_{\bs p}$ with the following one-particle energy:
\beqn
\omega_{\bs p} = \sqrt{{\bs k}^2 + (\sigma_1 + i \sigma_2)^2 + (\phi_1 + i\phi_2)^2}.
\label{eq:omega:0}
\eeqn

We use the axial (chiral) rotation~\eq{eq:rotation} applied to the combination $\sigma + i \gamma^5 \phi$ of now-complex fields $\sigma = \sigma_1 + i \sigma_2$ and $\phi = \phi_1 + i\phi_2$ to reduce an unphysical degree of freedom. Given the fact that the unitarity of the chiral transformation requires the parameter $\omega_A$ of Eq.~\eq{eq:rotation} to be real, we may get rid of only one real-valued field.  We choose to remove, following the standard approach to the NJL mode, the real component of the complex $\phi$ field, $\phi_1 = 0$.

Similarly to the case of the standard Hermitian theory, we take $\sigma_1 \neq 0$. In this case, the stability of the ground state of the theory requires $\sigma_2 = 0$. The requirements $\phi_1 = 0$ and $\sigma_2 = 0$ are enforced by taking the corresponding constants to be zero, $G_{\phi1} \to 0$ and $G_{\sigma2} \to 0$, respectively. According to the algebraic conditions~\eq{eq:G:relations}, we are left with the two linearly-dependent coupling constants $G_{\sigma1} = G$ and $G_{\phi2} = - G$ and two one-component fields $\sigma_1 \equiv \sigma$ and $\phi_2 \equiv - \phi_5$. Here we have redefined the fields again for the sake of convenience. 

Using the new non-Hermitian bosonization, we arrive to the following representation of the NJL model:
\beqn
\cL_{\NJL} = \frac{\phi_5^2 - (\sigma - m_0)^2}{2 G} 
+ \bar{\psi} \left( i \slashed\partial - \sigma - \gamma^5 \phi_5 \right)\psi.\quad
\label{eq:L:NJL:NH:2}
\eeqn
We immediately recognize that the field $\phi_5$ plays the role of the non-Hermitian fermionic mass $m_5$ already discussed in the context of the free theory~\eq{eq:cL} while the field $\sigma$ appears in the standard role of the usual Hermitian mass $m$. It is important to mention here that the effective NH-NJL model in Eq.(\ref{eq:L:NJL:NH:2}) possesses the symmetries described in Sec. \ref{sec:free:fermions} concerning $\mathcal{PT}$, $U(1)_V$, and $U(1)_A$ symmetries. These symmetries will be maintained when integrating out the fermionic field and obtaining a completely bosonized version of the NH-NJL model.

The coupling between the fermions and the the auxiliary field $\phi_5$ in the partially bosonized NJL Lagrangian~\eq{eq:L:NJL:NH:2} has the form of the anti-Hermitian Yukawa coupling that has been studied recently~\cite{Alexandre:2020bet,Alexandre:2020tba} in the phenomenological context of sterile neutrinos. 

The energy spectrum of fermions in the partially bosonized NJL theory~\eq{eq:L:NJL:NH:2} is, now:
\beqn
\omega_{\bs p} = \sqrt{{\bs k}^2 + \sigma^2 -\phi_5^2},
\label{eq:omega:M}
\eeqn
which is typical for the fermions which have both Hermitian $\sigma$ and non-Hermitian $\phi_5$ fermionic masses~\eq{eq:M:0}.

After integrating out the fermionic fields in the non-Hermitian theory~\eq{eq:L:NJL:NH:2}, we arrive to the effective potential on the scalar and pseudo-scalar non-Hermitian fields:
\beqn
V_{\mathrm{NH}}(\sigma,\phi_5) = \frac{(\sigma - m_0)^2 - \phi_5^2}{2 G} 
+ \Lambda^4 v\left(\frac{\sigma^2 - \phi_5^2}{\Lambda^2}\right),
\label{eq:V:NH}
\eeqn
where the function $v(x)$ is given in Eq.~\eq{eq:small:v}.

One notices that the minimal realization of a scalar non-Hermitian model involves two scalar fields, one of them is a true scalar while another one is often chosen as a pseudoscalar~\cite{Alexandre:2017foi,Alexandre:2017erl,Alexandre:2020gah}. This property is also maintained by the effective bosonized theory~\eq{eq:V:NH}. The form of the bosonized effective model~\eq{eq:V:NH} is qualitatively different from the multi-scalar model considered so far in the literature~\cite{Alexandre:2018uol,Mannheim:2018dur,Fring:2019hue}.

It it worth comparing the effective potential for the non-Hermitian ground state~\eq{eq:V:NH} with the potential for the Hermitian state~\eq{eq:V}. The Hermitian potential does not show dissipatively unstable behaviour for any value of the mean field $\sigma$.  The non-Hermitian potential in Eq.(\ref{eq:V:NH}) is stable, that is, it displays at least one minimum, provided the $\PT$ symmetry is unbroken:
\beqn
\sigma^2 \geqslant \phi_5^2.
\eeqn
These considerations could naively (and, incorrectly) imply that there is no obvious effect of the non-Hermiticity on the physical properties of the model. Indeed, we are free to denote the field combination in Eq.~\eq{eq:V:NH} as
\beqn
M^2 = \sigma^2 - \phi_5^2\geqslant 0,
\label{eq:M}
\eeqn
and to come back to the Hermitian model in terms of the new field $\sigma \equiv M$ with the same fermionic spectrum~\eq{eq:omega:M}: $\omega_{\bs p} = \sqrt{{\bs k}^2 + M^2}$.

Contrary to the Hermitian bosonization of the NJL model, there is no obvious interpretation of the real-valued field $\phi_5$ in terms of a fermionic condensate. For example, a variation of the action associated with the Lagrangian~\eq{eq:L:NJL:NH:2} with respect to the field $\phi_5$ would not give us an anticipated relation between the field $\phi_5$ and a fermionic bilinear at a saddle-point of the effective theory because
\beqn
\avr{\phi_5} \neq G \avr{{\bar \psi \gamma^5 \psi}}.
\label{eq:phi}
\eeqn
Indeed, the field $\phi_5$ at the left-hand side of Eq.~\eq{eq:phi} is a real-valued quantity while the fermionic condensate at the right-hand side of the same equation either vanishes or takes a purely imaginary value.

In an attempt to interpret the field $\phi_5$ in line of the fermionic condensate~\eq{eq:phi}, one could consider a possibility that the saddle point~\eq{eq:phi} is realized for a purely imaginary field $\phi_5$. In this case, we could redefine the field $\phi_5 = i \phi$ and come back to the Hermitian theory, in which the mean field $\phi$ may be removed further by the chiral rotation~\eq{eq:rotation}. This approach could be sustainable if the minimum of the potential at the purely imaginary $\phi_5$ is lower compared to its minimum at the real $\phi_5$. We will explore this possibility below.

Before finishing this section, we would like to comment that we could have arrived to the same non-Hermitian model~\eq{eq:L:NJL:NH:2} by employing only the first and the last of the identities of the alternative set~\eq{eq:identity:II} instead of the original identities~\eq{eq:identity:I}. This would be a fully legitimate operation. We could also naively use a non-Unitary chiral rotation~\eq{eq:rotation} with a purely imaginary $\omega_A$ and also come to the same conclusion. The latter operation is, however, logically forbidden because the transformation~\eq{eq:U1:A} with a purely imaginary $\omega_A$ is not a symmetry of the original theory.

\subsection{Non-Hermitian ground state in the chiral limit}

The non-Hermitian ground state of the system is determined by the minimum of the non-Hermitian potential~\eq{eq:V:NH} with respect to the fields $\sigma$ and $\phi_5$. Indeed, the potential determines the statistical weight of the mean-field configuration similarly as it happens in the Hermitian case which was discussed at the end of section~\eq{sec:NJL:H}. However, in the non-Hermitian ground state, there is one obvious subtlety: the potential $V_{\mathrm{NH}}$ may take a complex value if the $\PT$ symmetry is broken. Therefore, the statistical interpretation does not apply to the $\PT$-broken regime which should be unstable in any case. Below, we concentrate on the $\PT$ symmetric ground state.

We start this section by noticing that for $m_0=0$, 
the NJL potential in the bosonized mean-field representation~\eq{eq:V:NH} possesses the the emergent $U(1)$ symmetry group,
\beqn
U(1)_{\NH}:\ \ \left(\begin{array}{c}
\!\sigma\!\! \\[1mm]
\!\phi_5\!\!
\end{array}
\right)
\to
\left(\begin{array}{cc}
\cosh \omega_{\NH} & \sinh  \omega_\NH\\[1mm]
\sinh  \omega_\NH & \cosh  \omega_\NH
\end{array}
\right)
\left(\begin{array}{c}
\!\sigma\!\! \\[1mm]
\!\phi_5\!\!
\end{array}
\right)\!\!, \quad\
\label{eq:U1:NH}
\eeqn
which is parameterized by an arbitrary real-valued angle $ \omega_\NH \in \R$. The non-Unitary transformation~\eq{eq:U1:NH} corresponds a noncompact Abelian group which keeps invariant the combination of the scalar fields $\sigma^2 - \phi_5^2$.

In presence of condensates of $\sigma$ and/or $\phi_5$ fields, the non-Hermitian symmetry~\eq{eq:U1:NH} is broken spontaneously. According to the standard lore, this breaking should lead to new Goldstone bosons in the spectrum. The Goldstone bosons in the context of non-Hermitian field theories have been discussed in Refs.~\cite{Fring:2019xgw,Fring:2019hue} for both Abelian and non-Abelian theories. Here we firstly discuss the theory with $m_0=0$, and in the next subsection we analyze the effects of a nonzero current mass. 
 
In Fig.~\ref{fig:potentials}(a) we show the mean-field potential~\eq{eq:V:NH} in the chirally symmetric ($G < G_c$) phase. If we restrict ourselves to the purely Hermitian case, $\phi_5 = 0$, then the minimum of the potential is reached at the symmetric vacuum $\sigma = 0$ (marked by the small red sphere in the figure.) In the full non-Hermitian plane, the minimum is reached exactly at the border of the $\PT$ symmetric region, $\sigma = \pm \phi_5$. This vacuum state breaks the non-Hermitian symmetry group~\eq{eq:U1:NH}. 

Figure~\ref{fig:potentials}(b) shows the potential~\eq{eq:V:NH} in the chirally broken ($G > G_c$) phase. The minimum of the potential, shown by the blue line, respects the non-Hermitian symmetry~\eq{eq:U1:NH}. The green line shows the profile of the potential in the Hermiticity region $\phi_5 = 0$ which is denoted by the greenish plane. Both lines cross at the Hermitian minimum of the potential: $(\sigma,\phi_5)_H = (\sigma_{\mathrm{min}},0)$. The Hermitian minimum is related to all other non-Hermitian minima via a non-compact transformation~\eq{eq:U1:NH}.

Thus, we conclude that in the case of $m_0 = 0$, the non-Hermitian ground state with nonvanishing scalar Hermitian field $\sigma$ and the pseudoscalar non-Hermitian $\phi_5$ field can well be realized in the Hermitian NJL model. The non-Hermitian ground state breaks the non-compact non-Hermitian group~\eq{eq:U1:NH}.

\subsection{The instability associated to a non-zero $m_0$\label{sec:finitem0}}

\begin{figure*}[!tb]
\begin{center}
\includegraphics[width=1.5\columnwidth]{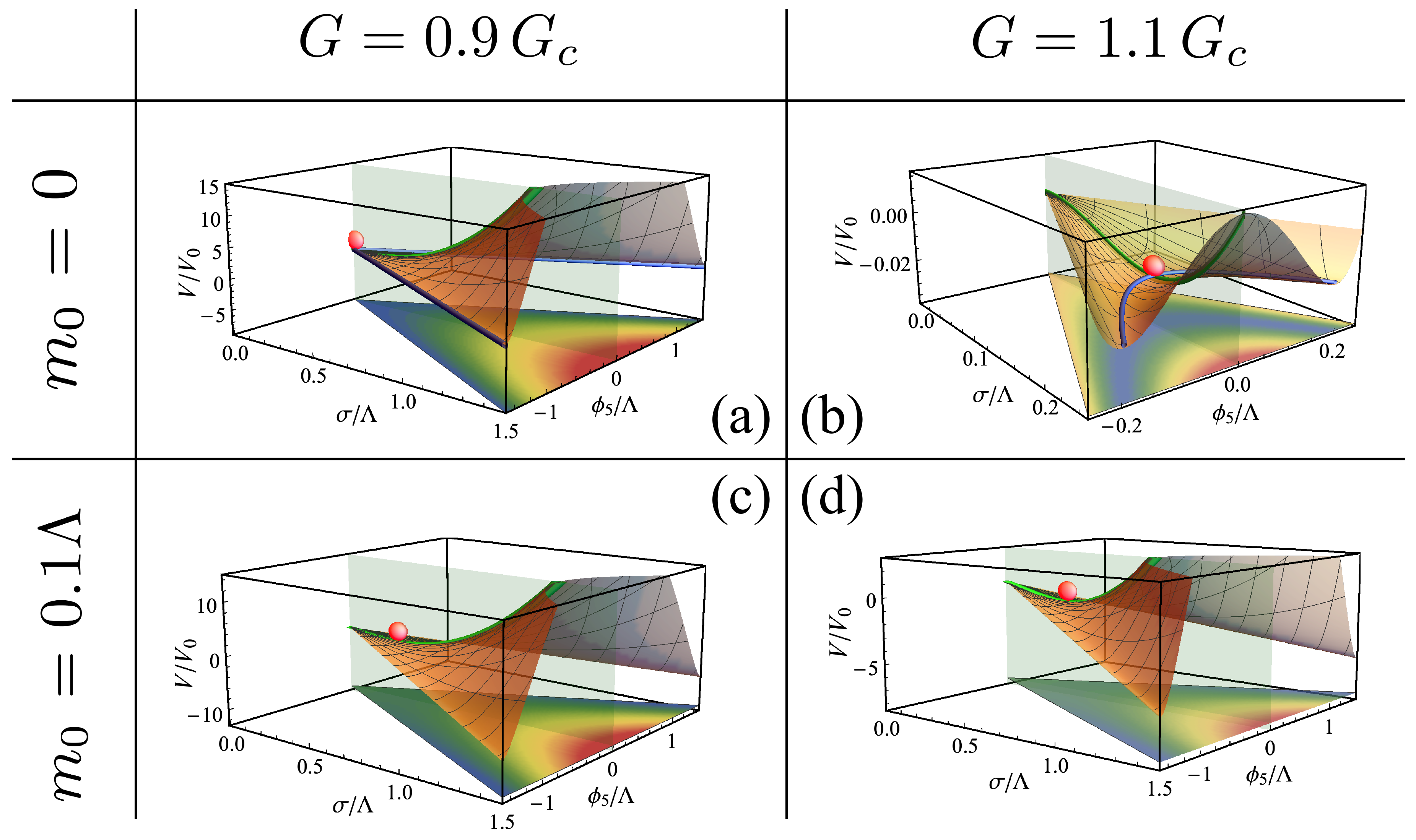}
\end{center}
\vskip -3mm
\caption{The mean-field potential~\eq{eq:V:NH} as the function of the {\it uniform, coordinate-independent} $\sigma$ and $\phi$ condensates in the chirally symmetric ($G < G_c$) and chirally broken ($G > G_c$) phases at zero ($m_0 = 0$) and non-zero ($m_0 = 0.1 \Lambda$) bare masses $m_0$. The critical coupling $G_c$ is given in Eq.~\eq{eq:Gc} and $V_0 = (\Lambda/2\pi)^4$. The projection on the $(\sigma,\phi)$ plane illustrates the region where the $\PT$ symmetry is unbroken. The density plot at the projection shows the values of the potential ranging from the low (the blue) to the high (the red) values. The greenish cross-section shows the Hermitian region ($\phi_5=0$) with the value of the potential highlighted by the green line and the minimum of the effective potential marked by the red spheres.
The blue line shows the degenerate minimum of the potential in the full $(\sigma,\phi)$ plane.}
\label{fig:potentials}
\end{figure*}

The presence of the mass $m_0$ in the potential~\eq{eq:V:NH} breaks, mildly and explicitly, the non-Hermitian symmetry~\eq{eq:U1:NH}. However, this small explicit symmetry breaking -- which would be non-harmful in the usual Hermitian theory, leads to the instability of the non-Hermitian theory. This fact can be observed both for the weakly coupled phase with $G<G_c$, Fig.~\ref{fig:potentials}(c), and for strongly coupled phase ($G > G_c$), Fig.~\ref{fig:potentials}(d).

The same conclusions may also be easily reached analytically in the case of a large condensate, $\sigma \gg m_0$. Then the vacuum solution follows two lines $\phi_5 \simeq \pm \sigma$ and the effective potential~\eq{eq:V:NH} is a linear function of the condensate:
\beqn
V(\sigma, \phi_5 = \pm \sigma) \simeq - \frac{m_0 \sigma}{G} + \mathrm{const}, \quad (\mbox{at\ } \sigma \gg m_0)
\label{eq:V:runaway}
\eeqn
which may be made arbitrarily low. While we do not investigate this issue in detail in the present study, we would like to comment that this instability might be removed by inhomogeneous condensates by bringing further terms to the free energy that depend on space derivatives. The inhomogeneous ground state appears naturally in the interacting fermionic systems such as the large--$N$ Gross-Neveu model (the NJL model in one spatial dimension) at finite temperature and finite chemical potential. This ground state is closely related to inhomogeneous superconductors in the Larkin-Ovchinnikov-Fulde-Ferrell phase~\cite{Schon:2000he,Thies:2006ti,Basar:2009fg}.

We conclude that the NJL does not possess a stable vacuum in terms of uniform condensates in the presences of a small explicit symmetry breaking given by small (in fact, an infinitely small) mass $m_0 \neq 0$ if the non-Hermitian condensate $\phi_5$ is allowed.

\section{Ginzburg-Landau analysis\label{sec:GLanal}}
\label{sec:GL}

In this section we use a Ginzburg-Landau (GL) effective potential to  analyze the ground state of the model. In previous sections we have focused on the possibility of NH constant values of the condensates. However, we did not provided any physical reason why this homogeneous state should be the actual ground state of the theory. As mentioned at the end of Sec.~\ref{eq:NJL:NH}, a constant field solution of the NH will be energetically unstable in the PT-broken sector, and there is the possibility of non-homogeneous solutions of the mean field theory. The analysis of the GL potential is the appropriate tool to investigate such possibility.

This expansion is known to work well only in proximity of a 
second-order phase transition: In doing this analysis, we assume that
the coupling constant is slightly larger the critical coupling so that 
the condensate is not large. 

We do not specify any concrete temperature or chemical potential here,
while we assume that the sixth order coefficients (defined in the next paragraphs) are positive to have
a bounded-from-below effective potential.
The structure of the GL potential, in particular the
relations between the coefficients of the polynomials and derivative
terms, will be adapted from that of
chiral symmetry breaking of the NJL model~\cite{Nickel:2009ke}.
The GL potential has to be consistent with the symmetry of the model,
in particular it should be invariant under the global transformation
in Eq.~\eqref{eq:U1:NH}. Besides, the GL potential, as a thermodynamic quantity, does not depend on any specific choice of metric operator that renders the theory consistent. In other words, after integrating out fermions in Eq.(\ref{eq:L:NJL:NH:2}) the resulting GL potential will not depend on this quantity~\cite{Brody:2016}.

Up to the sixth order in the fields
and fourth order in derivatives,
the GL potential for the model at hand has to have the form
\begin{eqnarray}
F_{\NH} &=& \frac{\alpha_2}{2}\vec\chi\cdot\vec\chi +   
\frac{\alpha_4}{4}(\vec\chi\cdot\vec\chi)^2 +
\frac{\alpha_6}{6}(\vec\chi\cdot\vec\chi)^3 \nonumber \\
&& +\frac{\beta_4}{4} {\bs \nabla} {\vec \chi}.{\bs \nabla} {\vec \chi} 
\nonumber \\
&&+\frac{\gamma_6}{6}({\vec \chi}.{\vec \chi}) \bigl({\bs \nabla} {\vec \chi}.{\bs \nabla} {\vec \chi}\bigr)^2+ \frac{\delta_6}{6}\left(\vec\chi\cdot\bm\nabla\vec\chi\right)^2\nonumber\\
&&+\frac{\varepsilon_6}{6}\Delta {\vec \chi}.\Delta {\vec \chi},
\label{eq:generalPPP}
\end{eqnarray}
where ${\vec \chi} = (\sigma,\phi_5)$ is the vector in the space of fields equipped with the hyperbolic metric: ${\vec \chi}.{\vec \chi} = \sigma^2 - \phi_5^2$ and $\bm\nabla{\vec \chi} \cdot\bm\nabla{\vec \chi} = (\bm\nabla\sigma)^2-(\bm\nabla\phi_5)^2$.
As mentioned earlier, stability requires that the sixth order coefficients are positive: $\alpha_6,\beta_6,\gamma_6,\delta_6,\varepsilon_6 > 0$.

The symmetry alone does not allow to fix the relations between the
coefficients of the expansion in Eq.~\eqref{eq:generalPPP}.
In order to simplify the GL potential
we use the known results of the NJL model with chiral 
condensate:
in fact, the GL potential for the non-Hermitian
model has to have the same invariance under
the chiral rotation of the simpler NJL model, 
augmented with invariance under rotation Eq.~\eqref{eq:U1:NH} 
and the latter can be implemented by the 
replacement $\sigma^2\rightarrow\sigma^2 - \phi_5^2$ in the potential
of the NJL model.
 
In the NJL model with a chiral condensate the free energy density
in the GL approximation is given by~\cite{Nickel:2009ke}
\beqn
& & F_\mathrm{NJL}(\sigma({\bs x})) = 
\frac{\alpha_2}{2} \sigma^2({\bs x})
+
\frac{\alpha_4}{4}
\bigl[\sigma^4({\bs x}) + ({\bs \nabla} \sigma({\bs x}))^2\bigr] 
\nonumber\\
& & \qquad + \frac{\alpha_6}{6} \left( \sigma^6({\bs x}) +
5\sigma^2({\bs x}) \bigl[{\bs \nabla} \sigma({\bs x})\bigr]^2 
+ \frac{1}{2}\bigl[\Delta \sigma({\bs x})\bigr]^2
\right),\nonumber\\
&&
\label{eq:NJL:scalar}
\eeqn
The effective potential for the non-Hermitian ground state 
can be obtained from Eq.~\eqref{eq:NJL:scalar} 
by the replacements
\begin{eqnarray}
&&\sigma^2\rightarrow \vec\chi\cdot\vec\chi, \\
&&\left(\bm\nabla\sigma\right)^2\rightarrow \bm\nabla\vec\chi\cdot
\bm\nabla\vec\chi,\\
&&\sigma^2\left(\bm\nabla\sigma\right)^2
\rightarrow
\left(\vec\chi\cdot\vec\chi\right)
\bm\nabla\vec\chi\cdot
\bm\nabla\vec\chi~~\mathrm{or}~~\left(\vec\chi\cdot\bm\nabla\vec\chi\right)^2,\label{eq:ambi:a1}\\
&&\left(\Delta\sigma\right)^2\rightarrow
\Delta\vec\chi\cdot\Delta\vec\chi,
\end{eqnarray}
where ${\vec \chi} = (\sigma,\phi_5)$ has been defined above.
Notice that the NJL invariant on the left-hand side
of Eq.~\eqref{eq:ambi:a1} leads to two possible terms in the
non-Hermitian model, namely those with coefficients 
$\gamma_6$ and $\delta_6$ in Eq.~\eqref{eq:generalPPP}.
We analyze the two limiting possibilities here, namely $\gamma_6=0$ and
$\delta_6=0$, noticing that in the former case we get a stable GL 
potential while in the latter case the free energy is unbounded from
below.

Starting with $\delta_6=0$,
the GL free energy of the non-Hermitian model reads
\beqn
& & F_{\NH}({\vec \chi}({\bs x})) =
\frac{\alpha_2}{2} {\vec \chi}.{\vec \chi}
\label{eq:NJL:scalar:2} \\
& & 
\qquad + \frac{\alpha_4}{4} \Bigl[({\vec \chi}.{\vec \chi})^2 + {\bs \nabla} {\vec \chi}.{\bs \nabla} {\vec \chi}
\Bigr] 
\nonumber \\
& & \qquad + \frac{\alpha_6}{6} \Bigl[ \bigl({\vec \chi}.{\vec \chi}\bigr)^3 + 
5({\vec \chi}.{\vec \chi}) \bigl({\bs \nabla} {\vec \chi}.{\bs \nabla} {\vec \chi}\bigr)^2
+ \frac{1}{2} \Delta {\vec \chi}.\Delta {\vec \chi}
\Bigr]. \nonumber
\eeqn

It is now useful to introduce two new fields, $\xi$ and $\theta$, by means of the following transformation
\beqn
\sigma=\xi \cosh\theta, \qquad \phi_5=\xi \sinh\theta.
\label{eq:Delta:rotation}
\eeqn
Using the parameterization~\eq{eq:Delta:rotation}, we rewrite the free energy~\eq{eq:NJL:scalar:2} as follows:
\beqn
& & F_{\NH}(\xi, \theta) =
\frac{\alpha_2}{2} \xi^2 +
\frac{\alpha_4}{4} \xi^4 + \frac{\alpha_6}{6} \xi^6
\nonumber \\
& & 
\qquad + \frac{\alpha_4}{4} \Bigl[({\bs \nabla} \xi)^2 - \xi^2 ({\bs \nabla} \theta)^2 \Bigr] 
\nonumber\\
& & \qquad + \frac{5\alpha_6}{6} \Bigl[
\xi^2 ({\bs \nabla} \xi)^2 - \xi^4 ({\bs \nabla} \theta)^2 \Bigr] \nonumber \\
& & \qquad + \frac{\alpha_6}{12} \Bigl[ \Delta \xi + \xi ({\bs \nabla} \theta)^2\bigr]^2
\nonumber \\
& & \qquad - \frac{\alpha_6}{12}  \bigl[2 ({\bs \nabla} \xi \cdot {\bs \nabla} \theta) + \xi \Delta \theta \bigr]^2,
\label{eq:NJL:scalar:3}
\eeqn
where we used the identities:
\beqn
{\vec \chi}.{\vec \chi} & = & \xi^2, \\
{\bs \nabla} {\vec \chi}.{\bs \nabla} {\vec \chi} & = & ({\bs \nabla} \xi)^2 - \xi^2 
({\bs \nabla} \theta)^2, \\
\Delta {\vec \chi}.\Delta {\vec \chi} & = & 
\bigl[\Delta \xi + \xi ({\bs \nabla} \theta)^2\bigr]^2 
- \bigl[2 ({\bs \nabla} \xi \cdot {\bs \nabla} \theta) + \xi \Delta \theta\bigr]^2.\qquad
\eeqn

The free energy~\eq{eq:NJL:scalar:3} is invariant under rigid shifts of the hyperbolic field $\theta \to \theta + \theta_0$. This freedom corresponds to the transformations of the global non-Hermitian group~\eq{eq:U1:NH}. The corresponding non-Hermitian Nambu-Goldstone boson is represented by the single "pion" field $\phi_5$, which, in the linear order, is realized as the hyperbolic phase $\theta$ with the kinetic-only action. Notice that the kinetic action enters the free energy with the negative sign which is the expected property in the non-Hermitian scalar gauge theories~\cite{Alexandre:2017foi}. The negative sign in front of the kinetic term of the $\theta$ field is the reason why the presence of the $\xi \neq 0$ condensate breaks the translational symmetry of the strongly-coupled phase. We discuss this question below.

For $\alpha_4>0$, the terms proportional to $(\bm\nabla\xi)^2$
and $\xi^2(\bm\nabla\theta)^2$ in~\eqref{eq:NJL:scalar:3}
respectively increase and decrease the free energy
for an inhomogeneous ground state: 
therefore, for $\alpha_4>0$ 
it is likely that the lowest free energy is achieved
by a ground state of the form
\begin{equation}
   \xi = \xi_0, \qquad \theta = \theta(x), 
\label{eq:vacuum:general} 
\end{equation}
where $\xi_0$ is a constant. For simplicity, we limit ourselves here to
the following ansatz:
\beqn
\xi = \xi_0, \qquad \theta = \theta_0 + {\bs k}\cdot{\bs x}, 
\label{eq:vacuum}
\eeqn
where ${\bs k}$ is a constant wave vector and $\theta_0$ 
is an arbitrary constant phase. 
The vacuum with a nonzero wavevector ${\bs k}$ breaks the rotational symmetry of the ground state.

Using the ansatz~\eqref{eq:vacuum} in Eq.~\eqref{eq:NJL:scalar:3}
we get the free energy of the ground state:
\beqn
F_{\NH}{\biggl|}_{\xi = \xi_0; \theta = {\bs k} {\bs x}} & = &
\frac{\alpha_2}{2} \xi^2_0 +
\frac{\alpha_4}{4}  \xi^4_0   + \frac{\alpha_6}{6} \xi^6_0
\nonumber \\
& & \hskip -5mm
- \biggl(\frac{\alpha_4}{4}
+ \frac{5\alpha_6}{6} \xi^2_0 \biggr) \xi^2_0 {\bs k}^2 + \frac{\alpha_6}{12} \xi_0^2 {\bs k}^4;
\label{eq:NJL:scalar:4}
\eeqn
the above equation tells us that an inhomogeneous ground state with a non-zero wavevector $\bm k\neq 0$ lowers the free energy
of the system. 

The direction of $\bm k$ in the ground state will be
chosen spontaneously by the system as it does not affect
the free energy, while 
the magnitude of $\bm k$, namely $k_0$,
minimizes $F_{\NH}$. Explicitly, the length of the ground-state wavevector is
\beqn
k_0 = \sqrt{\frac{3 \alpha_4 + 10 \alpha_6\xi_0^2}{2\alpha_6}}.
\label{eq:k:vac}
\eeqn
Using this expression for $k_0$ in Eq.~\eqref{eq:NJL:scalar:4}, we
get the free energy at the minimum:
\beqn
F_{\NH}{\biggl|}_{\xi = \xi_0; {\bs k} = {\bs k}_0} & = &
\frac{\tilde\alpha_2}{2}\xi_0^2 + \frac{\tilde\alpha_4}{4}\xi_0^4
+\frac{\tilde\alpha_6}{6}\xi_0^6,\label{eq:ren:a1}
\eeqn
with
\begin{eqnarray}
\tilde\alpha_2 &=& \alpha_2 -\frac{3\alpha_4^2}{8\alpha_6},\label{eq:a1}\\
\tilde\alpha_4 &=& \alpha_4 -5\alpha_4,\label{eq:a2}\\
\tilde\alpha_6 &=& \alpha_6 - \frac{25\alpha_6}{2}.\label{eq:a3}
\end{eqnarray}
The free energy difference between the states with inhomogeneous and
homogeneous condensates is
\begin{equation}
    \Delta F = -\frac{3\alpha_4^2}{16\alpha_6}\xi_0^2
    -\frac{5\alpha_4}{4}\xi_0^4 -\frac{25\alpha_6}{12}\xi_0^6.
    \label{eq:deltaF:abc}
\end{equation}
Notice that the sixth order term in Eq.~\eq{eq:deltaF:abc} is negative,
which makes the free energy unbounded from below.
It is likely that in this case a resummation of all orders
in the condensate is necessary. We thus conclude that 
the GL analysis alone is not enough to determine completely the ground state,
although Eq.~\eqref{eq:deltaF:abc} seems to point in the direction of a free energy  is less favorable if the inhomogeneous ground state is taken.

Next we turn to the choice $\gamma_6=0$.
In this case
we replace 
\begin{equation}
    \xi^2(\bm\nabla\xi)^2-\xi^4(\bm\nabla\theta)^2
    \rightarrow
     \xi^2(\bm\nabla\xi)^2\label{eq:ambi:a3}
\end{equation}
in the third line of Eq.~\eqref{eq:NJL:scalar:3}, and
instead of Eq.~\eqref{eq:NJL:scalar:4} we would have 
\beqn
F_{\NH}{\biggl|}_{\xi = \xi_0; \theta = {\bs k} {\bs x}} & = &
\frac{\alpha_2}{2} \xi^2_0 +
\frac{\alpha_4}{4}  \xi^4_0   + \frac{\alpha_6}{6} \xi^6_0
\nonumber \\
& & \hskip -5mm
-  \frac{\alpha_4}{4}
 \xi^2_0 {\bs k}^2 + \frac{\alpha_6}{12} \xi_0^2 {\bs k}^4;
\label{eq:NJL:scalar:4bis}
\eeqn
this would give
\beqn
k_0 = \sqrt{\frac{3 \alpha_4}{2\alpha_6}}.
\label{eq:k:vacbis}
\eeqn
instead of Eq.~\eqref{eq:k:vacbis}, and
\begin{eqnarray}
\tilde\alpha_2 &=& \alpha_2 -\frac{3\alpha_4^2}{8\alpha_6},\\
\tilde\alpha_4 &=& \alpha_4,\\
\tilde\alpha_6 &=& \alpha_6.
\end{eqnarray}
instead of Eqs.~\eqref{eq:a1}-\eqref{eq:a3}.
In this case, the sign of the sixth order term coincides
with that of $\alpha_6$ which is assumed to be positive,
therefore the free energy is bounded from below and
the GL analysis can be completed by estimating the loss in free energy
due to the inhomogeneous condensate,
\begin{equation}
    \Delta F = -\frac{3\alpha_4^2}{16\alpha_6}\xi_0^2.
    \label{eq:deltaF:def}
\end{equation}
The numerical value of $\Delta F$ depends on the details of the
microscopic model; however, regardless of these details
we can conclude that 
the inhomogeneous condensation lowers the free energy of the system.

\section{Outlook}

The generalization of our approach to a system with $N_f > 1$ flavors (species) of fermions is rather straightforward. Similarly to the standard Hermitian NJL model~\cite{Klevansky:1992qe}, the fermions of different flavors form $N^2_f - 1$ pseudoscalar fields $\phi^a_5$ which couple with the $N_f$-multiplet of the fermionic field $\Psi = (\psi_1, \dots, \psi_{N_f})^T$ via the non-Hermitian Yukawa coupling:
\beqn
\cL_{\NJL} = \frac{{\bs \phi}_5^2 - (\sigma - m_0)^2}{2 G} 
+ \bar{\Psi} \left( i \slashed\partial - \sigma \bbbone - \gamma^5 \lambda^a  \phi^a_5 \right)\Psi.\qquad
\label{eq:L:NJL:NH:Nf}
\eeqn
Here $\lambda^a$ are the generators of the $SU(N_f^2 -1)$ flavor group and $\sigma$ is the scalar Hermitian field. The integration of over the fermionic degrees of freedom $\Psi$ produces the one loop action with the pseudoscalar iso-vector field ${\bs \phi}_5 = (\phi^1_5, \dots \phi^{{}_{N_f^2-1}}_5)^T$. In the uniform vacuum, it is sufficient to replace $\phi_5^2 \to {\bs \phi}^2_5$ in the effective action~\eq{eq:V:NH}. The effective bosonic action respects a non-compact symmetry group $U(1,N_f^2-1)$ which generalizes the single-flavor non-Hermitian group~\eq{eq:U1:NH} to the multiflavor case. Notice that while for a single-flavor fermion the non-Hermitian symmetry breaking leads to the appearance of the non-Hermitian mass gap given by the last term in Eq.~\eq{eq:L:NJL:NH:2}, in the multi-flavor case the non-Hermitian ground state may be traced to the emergence of the non-Hermitian Yukawa interaction in the last term of the semi-bosonized Lagrangian~\eq{eq:L:NJL:NH:Nf}.

One of the problems that might require future attention concerns the possible mitigation of the destabilizing effect of a finite current mass $m_0$ on the non-Hermitian ground state of the system as has been noted briefly in Section~\ref{sec:finitem0}. To elucidate this problem, let us follow the notation of Section~\ref{sec:GLanal} and parametrize, similarly to Eq.~\eq{eq:Delta:rotation}, the fields $\sigma+m_0=\xi \cosh\theta$ and $\phi_5=\xi \sinh\theta$. The partially bosonized theory gets now the following form:
\begin{eqnarray}
\mathcal{L} = -\frac{1}{2G}(\xi^2+m^2_0-2\xi m_0\cosh\theta) + \bar{\psi}(i\slashed{\partial}-\xi)\psi. \qquad
\end{eqnarray}
The first term in the effective theory gives us the tree-level potential for the bosonic fields $\xi$ and $\theta$:
\begin{eqnarray}
V_0(\xi,\theta) =\xi^2+m^2_0-2 m_0 \xi \cosh\theta,
\label{eq:V:0}
\end{eqnarray}
which is unbounded from below in the direction of the field $\theta$ if $m_0 \neq 0$. The problem is clearly absent in the chiral limit at $m_0 = 0$. 

The absence of a finite minimum of the tree-level potential $V_0$ at $m_0 \neq 0$ cannot be compensated by the fermionic determinant because the determinant may only add the one-loop term $v(\xi/\Lambda)$ to the potential~\eq{eq:V:0}. The one-loop correction, computed in Eq.(\ref{eq:small:v}), gives a contribution depending on $\xi$, and, consequently, it cannot lift out the instability of the classical potential~\eq{eq:V:0} in the direction of the field $\theta$. Therefore the quantum corrections can not solve the problem of the absence of the ground state in the minimal NJL model~\eq{eq:L:NJL:1} in the presence of the explicit chiral symmetry breaking, $m_0 \neq 0$.

\section{Conclusions}

We elaborated the physical consequences of a scenario in which the standard Hermitian Nambu--Jona-Lasinio model spontaneously develops a non-Hermitian $\PT$-symmetric ground state. Our scenario challenges the standard assumption that the ground state of a Hermitian theory is a Hermitian state. The non-Hermitian ground state is characterized by the presence of specific condensates that make the quasiparticle excitations non-Hermitian. In the context of the NJL model, the non-Hermitian ground state develops a complex axial condensate which gives a non-Hermitian contribution to the quark's mass. We call this vacuum ground state as a non-Hermitian ground state.

In the semi-bosonized model, the non-Hermitian ground state is catalyzed by the presence of a  dynamically generated non-Hermitian Yukawa coupling. The unbroken $\PT$ symmetry of the system guarantees that the spectrum is real. In the exact chiral limit, i.e. in the absence of the explicit chiral symmetry breaking in the NJL model, the uniform non-Hermitian ground state has the same (finite) free energy density as the usual Hermitian ground state. 

The known Hermitian and new non-Hermitian ground states are related to each other by a non-Unitary non-compact global symmetry which is spontaneously broken both in weak- and strong-coupling regimes of the model. In the chiral limit at strong coupling, the non-Hermitian ground state develops inhomogeneity, which breaks the translational symmetry of the state. At weak coupling, the ground state is a spatially uniform state, which lies at the boundary between the $\PT$-symmetric and $\PT$-broken phases. Outside the chiral limit, at a nonzero current mass $m_0 \neq 0$, the minimal NJL model does not possess a stable ground state because the free energy of the state is unbounded from below. The $m_0 \neq 0$ ground state may perhaps be stabilized in non-minimal NJL models with higher-order fermionic vertices.

\acknowledgments

M.~R. is grateful to John Petrucci for inspiration. 
M.~R. is supported by the National Science Foundation of China (Grants No.11805087 and No. 11875153)
and by the Fundamental Research Funds for the Central Universities (grant number 862946).
M.N.C. is partially supported by Grant No. 0657-2020-0015 of the Ministry of Science and Higher Education of Russia. A.C. acknowledges financial support through MINECO/AEI/FEDER, UE Grant No.
FIS2015-73454- JIN and European Union structural
funds, the Comunidad Aut\'onoma de Madrid (CAM)
NMAT2D-CM Program (S2018-NMT-4511), and the
Ram\'on y Cajal program through the grant RYC2018-
023938-I.

\appendix
\section{Gradient expansion in the effective action}

The bosonized NJL model after integrating out fermions gets the following form:
\begin{eqnarray}
\mathcal{L}_{NJL}&=&-\frac{1}{2G}(\sigma^{2}(x)-\phi^{2}_5(x))\nonumber\\
&+& i\,{\mathrm{Tr}}\, \log(G^{-1}_0-(\sigma+\gamma_5\phi_5)),
\end{eqnarray}
where  $G^{-1}_0=i\slashed{\partial}$. The trace operator, ``Tr'', includes for the matrix trace ``tr'' as well as an integral over spatial coordinates.

Factorizing $G^{-1}_0$ and expanding the logarithm in the above expression gives us:
\begin{eqnarray}
\mathcal{L}_{NJL}&=&-\frac{1}{2G}(\sigma^{2}(x)-\phi^{2}_5(x))+Tr[G^{-1}_0]\nonumber\\
&-&i\sum^{\infty}_{n=1}\frac{1}{n}Tr[G_0(\sigma+\gamma_5\phi_5)]^n.
\end{eqnarray}

Let us consider the case $n=2$:
\begin{eqnarray}
\Gamma_{n=2}&=&\frac{i}{2}\int dx dx\pr tr(G_0(x,x\pr)(\sigma(x\pr)+\gamma_5\phi_5(x\pr))\nonumber\\
&&G_0(x\pr,x)(\sigma(x)+\gamma_5\phi_5(x)))\nonumber\\
&=&\frac{i}{2}\int dx dx\pr tr[G_0(x,x\pr)G_0(x\pr,x)](\sigma(x\pr)\sigma(x)\nonumber\\
&-&\phi_5(x\pr)\phi_5(x)).
\end{eqnarray}

In order to perform the derivative expansion, it is mandatory to go to momentum representation, where the functional $\Gamma_{n=2}$ reads as follows:
\begin{eqnarray}
\Gamma_{n=2}=\int\frac{d^D q}{(2\pi)^D}\Pi_2(q)(\sigma(-q)\sigma(q)-\phi_5(-q)\phi_5(q)),
\end{eqnarray}
where
\beq
\Pi_2(q)=\frac{i}{2}\int \frac{d^D k}{(2\pi)^D}tr[G_0(k)G_0(k-q)],\label{eq:Pi2}
\eeq
and $G_0(k)=1/\gamma^{a}k_a$. With this, we can do the following:
\begin{eqnarray}
&&tr[G_0(k)G_0(k-q)]\simeq tr[G_0(k)G_0(k)]\nonumber\\
&+& tr[G_0(k)G_0(k)\frac{\partial G^{-1}_0}{\partial q_a}G_0(k)]q_a\nonumber\\
&+&tr[G_0(k)G_0(k)\frac{\partial G^{-1}_0}{\partial k_a}G_0(k)\frac{\partial G^{-1}_0}{\partial k_b}G_0(k)]q_a q_b,\label{eq:productexpansion}
\end{eqnarray}
where we have used the fact that $G^{-1}_0$ is a linear polynomial in the momenta. We plug the above expansion, $G_0=\gamma^{a}k_a/k^2$, and $\partial G^{-1}_0/\partial k_a=\gamma^{a}$ into Eq.~(\ref{eq:Pi2}). This trick corresponds to the derivative expansion of the effective action $\Gamma_{n=2}$. It is easy to see that the second term in (\ref{eq:productexpansion}) will be odd in the integration momentum, so it will not contribute to this expansion.

Now we can make the connection with the GL expansion with local terms as all the integrals are formally done at $q=0$. Then we have just to care of performing divergent integrals in the infrared (connection to non-renormalizable theories). Also, we need the following traces in $D=4$ ($q
^2=q\cdot q=q^{\mu}q_{\mu}$):
\beq
tr[\gamma^{a}\gamma^{b}]=4 \eta^{ab},
\eeq
\beq
I^{abcd}=tr[\gamma^{a}\gamma^{b}\gamma^{c}\gamma^{d}]=4(\eta^{ab}\eta^{cd}-\eta^{ac}\eta^{bd}+\eta^{ad}\eta^{bc}),
\eeq
\begin{eqnarray}
F^{abcdef}&=&tr[\gamma^{a}\gamma^{b}\gamma^{c}\gamma^{d}\gamma^{e}\gamma^{f}]=\eta^{ab}I^{cdef}-\eta^{ac}I^{bdef}
\nonumber\\
&+&\eta^{ad}I^{bcef}-\eta^{ae}I^{bcdf}+\eta^{af}I^{bcde}.
\end{eqnarray}

After a bit of algebra, we find
\begin{eqnarray}
tr[G_0(k)G_0(k-q)]&=&4\frac{1}{k^2}+8\frac{(k\cdot q)^2}{(k^2)^3} - 4\frac{q^2}{(k^2)^2}. \quad
\end{eqnarray}
Next, we rotate the integral in (\ref{eq:Pi2}) to the Euclidean space (there appears an extra $-i$ factor):
\beq
\Pi_2(q)=\frac{4}{2}\int \frac{d^4 k}{(2\pi)^4}\frac{1}{k^2}+\frac{8}{2}\int \frac{d^4 k}{(2\pi)^4}\frac{(k\cdot q)^2}{k^6}-\frac{4}{2}\frac{q^2}{k^4}.\label{eq:Pi2int}
\eeq
Performing the angular integration in four dimensions, and putting both UV and IR cutoffs ($\Lambda$ and $l$ respectively), we finally obtain:
\beq
\Pi_2(q)=\frac{\Lambda^2}{8\pi^2}-\frac{q^2}{8\pi^2}\log\left(\frac{\Lambda}{l}\right).
\eeq
From here we can read off $\alpha_2$ and $\alpha_4$ coefficients:
\begin{eqnarray}
\mathcal{S}_{0+2}&=&\int dx \frac{1}{2}\underbrace{(-\frac{1}{G}+\frac{\Lambda^2}{4\pi^2})}_{\alpha_2}(\sigma^2-\phi^2_5)-\nonumber\\
&-&\frac{1}{2}\underbrace{\frac{1}{4\pi^2}\log\left(\frac{\Lambda}{l}\right)}_{\alpha_4}((\nabla\sigma)^2-(\nabla\phi^2_5))
\end{eqnarray}

\begin{widetext}
Let us consider the $n=4$ contribution to the effective action (this time with the right sign), $\Gamma_{n=4}$:
\begin{eqnarray}
\Gamma_{4}&=&-\frac{i}{4}\int \Pi^{4}_{i=1}d x_i tr[(\sigma(x_1)-\gamma_5\phi_5(x_1))G_0(x_1-x_2)(\sigma(x_2)-\gamma_5\phi_5(x_2))G_0(x_2-x_3)\cdot\nonumber\\
&\cdot&(\sigma(x_3)-\gamma_5\phi_5(x_3))G_0(x_3-x_4)(\sigma(x_4)-\gamma_5\phi_5(x_4))G_0(x_4-x_1)].\label{eq:raweffaction41}
\end{eqnarray}
Out of the $16$ terms in the expression (\ref{eq:raweffaction41}) only $8$ will contribute to the expression, as they have an even number of $\gamma_5$ matrices in the product of propagators. Taking into account that the $\gamma_5$ matrices anticommute with the Dirac matrices $\gamma_{\mu}$, the effective action $\Gamma_{4}$ can be written as
\begin{eqnarray}
\Gamma_{4}&=&-\frac{i}{4}\int \Pi^{4}_{i=1}d x_i tr[G_0(x_1-x_2)G_0(x_2-x_3)G_0(x_3-x_4)G_0(x_4-x_1)]\cdot\nonumber\\
&\cdot&(\sigma(x_1)\sigma(x_2)\sigma(x_3)\sigma(x_4)-\sigma(x_1)\sigma(x_2)\phi_5(x_3)\phi_5(x_4)+\sigma(x_1)\phi_5(x_2)\sigma(x_3)\phi_5(x_4)\nonumber\\
&-&\sigma(x_1)\phi_5(x_2)\phi_5(x_3)\sigma(x_4)+\phi_5(x_1)\sigma(x_2)\phi_5(x_3)\sigma(x_4)-\phi_5(x_1)\sigma(x_2)\sigma(x_3)\phi_5(x_4)\nonumber\\
&-&\phi_5(x_1)\phi_5(x_2)\sigma(x_3)\sigma(x_4)+\phi_5(x_1)\phi_5(x_2)\phi_5(x_3)\phi_5(x_4)).
\label{eq:raweffaction42}
\end{eqnarray}
It will be convenient to write the expression (\ref{eq:raweffaction42}) in momentum space.
To avoid too long expressions, we will write up the first term in Eq.(\ref{eq:raweffaction42})($(dq)\equiv\frac{d^D q}{(2\pi)^D}$):
\begin{eqnarray}
\Gamma^{(1)}_4=\frac{1}{4}\int (dq)(dp)(dr)\Pi_4(q,p,r)\sigma(q)\sigma(p)\sigma(r)\sigma(-q-p-r),
\end{eqnarray}
with
\begin{eqnarray}
\Pi_4(q,p,r)=-i\int(dk)tr[G_0(k)G_0(k-p)G_0(k-p-r)G_0(k+q)],\label{eq:susceptibility4}
\end{eqnarray}
being the fourth order susceptibility. The remaining terms in Eq.(\ref{eq:raweffaction42}) are similar, just substituting the corresponding $\sigma$ fields by $\phi_5$ fields.

One can make the connection with the Ginzburg-Landau effective theory by approximating the function $\Pi_4(q,p,r)$ by its value at $q=p=r=0$, $\Pi_4(0,0,0)\equiv\alpha_4$, and then go back to real space. Again, using the term with four $\sigma$ fields we get: \begin{eqnarray}
\Gamma^{(1)}_4&=&\frac{\alpha_4}{4}\int (dq)(dp)(dr)\sigma(q)\sigma(p)\sigma(r)\sigma(-q-p-r)\nonumber\\
&=&\frac{\alpha_4}{4}\int(dp)(dq)(dr)\int\Pi^{4}_{i=1}e^{i x_1 q}e^{i x_2 p}e^{i x_3 r} e^{ix_4(-p-q-r)}\sigma(x_1)\sigma(x_2)\sigma(x_3)\sigma(x_4)\nonumber\\
&=&\frac{\alpha_4}{4}\int\Pi^{4}_{i=1}dx_i\delta(x_1-x_4)\delta(x_2-x_4)\delta(x_3-x_4)\sigma(x_1)\sigma(x_2)\sigma(x_3)\sigma(x_4)\nonumber\\
&=&\frac{\alpha_4}{4}\int dx \sigma^{4}(x).
\end{eqnarray}
Doing the same for the rest of the terms, and collecting all them in (\ref{eq:raweffaction42}) under this approximation, we have
\begin{eqnarray}
\Gamma_4&=&\frac{\alpha_4}{4}\int dx (\sigma^4(x)-\sigma^2(x)\phi^2_5(x)+\sigma^2(x)\phi^2_5(x)-\sigma^2(x)\phi^2_5(x)+\sigma^2(x)\phi^2_5(x) \nonumber\\
&-&\sigma^2(x)\phi^2_5(x)-\sigma^2(x)\phi^2_5(x)+\phi^4_5(x))=\frac{\alpha_4}{4}\int dx(\sigma^2(x)-\phi^2_5(x))^2\nonumber\\
&=&\frac{\alpha_4}{4}\int dx (\vec{\chi}\cdot\vec{\chi})^2.
\end{eqnarray}

This expression coincides with the free energy $F_{\NH}({\vec \chi})$ of Eq.~\eq{eq:generalPPP} when the latter is written in terms of $\vec{\chi}$. Also, the expression (\ref{eq:susceptibility4}) is the starting point to start the derivative expansion to get the terms proportional to $\alpha_6$.
\end{widetext}


%

\end{document}